\documentclass[twocolumn]{revtex4}

\usepackage{graphicx}
\usepackage{dcolumn}
\usepackage{amsmath}
\usepackage{amssymb}

\newcommand{\dotprod}{{\scriptscriptstyle \stackrel{\bullet}{{}}}}

\begin{document}

\title{The Stochastic Dynamics of an Array of Atomic Force Microscopes in a Viscous Fluid}

\author{M. T. Clark}
 \email{clarkmt@vt.edu}
\author{M. R. Paul}
\affiliation{Department of Mechanical Engineering, Virginia
Polytechnic and State University, Blacksburg, Virginia 24061}

\date{\today}

\begin{abstract}
We consider the stochastic dynamics of an array of two closely
spaced atomic force microscope cantilevers in a viscous fluid for
use as a possible biomolecule sensor. The cantilevers are not
driven externally, as is common in applications of atomic force
microscopy, and we explore the stochastic cantilever dynamics due
to the constant buffeting of fluid particles by Brownian motion.
The stochastic dynamics of two adjacent cantilevers are correlated
due to long range effects of the viscous fluid. Using a recently
proposed thermodynamic approach the hydrodynamic correlations are
quantified for precise experimental conditions through
deterministic numerical simulations. Results are presented for an
array of two readily available atomic force microscope
cantilevers. It is shown that the force on a cantilever due to the
fluid correlations with an adjacent cantilever is more than 3
times smaller than the Brownian force on an individual cantilever.
Our results indicate that measurements of the correlations in the
displacement of an array of atomic force microscopes can detect
piconewton forces with microsecond time resolution.
\end{abstract}


\maketitle

\section{Introduction}

The advent of micro and nanotechnology has ushered forth
measurement techniques with unprecedented
sensitivity~\cite{bustamante:2000,roukes:2000:1,craighead:2000,arlett:2006}.
For example the atomic force microscope (AFM)~\cite{binnig:1986}
has revolutionized surface science by enabling topographical
measurements with atomic
precision~\cite{garcia:2002,giessibl:2003,hansma:1994}. Despite
such major advances, a difficult challenge that remains is the
measurement of the real time dynamics of single molecules in their
natural aqueous environments. For example, the dynamics of
proteins such as conformational changes or substrate metabolysis
occur on force scales of 10's of piconewtons and time scales of
10's of milliseconds~\cite{radmacher:1994,thomson:1996}. This
parameter regime is difficult to reach using current technologies.
A promising approach is the use of micro and nanoscale
cantilevers~\cite{viani:1999,paul:2004,arlett:2005}.

Atomic force microscopy relies upon detecting and interpreting the
dynamics of an externally driven micron scale cantilever as it
interacts with a sample. The standard approaches of contact,
tapping, and non-contact mode involve the active control and
driving of the cantilever probe to interact with the sample of
interest. The ultimate sensitivity of these measurements is
limited by the underlying stochastic thermal motion of the
cantilever. An alternative measurement approach is to exploit
these stochastic fluctuations. This can be accomplished by placing
a passive (or undriven) cantilever in fluid and measuring the
resulting stochastic dynamics. The measurement of the thermal
noise spectrum is a commonly used calibration technique in atomic
force microscopy~\cite{martin:1987,cleveland:1993}.

Using a passive detection technique, successful measurement then
relies upon detecting and interpreting the change in the
stochastic dynamics of the cantilever due to the presence or the
dynamics of the sample. For example, a biomolecule could be
tethered between a cantilever and a surface. The stochastic
dynamics of the cantilever would then be altered by the dynamics
of the linking molecule and its interaction with the fluid as well
as the proximity of the surface upon which the molecule is
attached. The force sensitivity of this measurement is limited by
the Brownian force on the cantilever in the absence of the target
biomolecule, i.e. if the dynamics of the attached protein induce
forces on the cantilever less than the Brownian force on the
cantilever the protein dynamics will be extremely difficult to
detect. For a typical atomic force microscope the Brownian force
on a single cantilever can be on the order of 100pN (discussed
further later) which is too large for many interesting biological
measurements~\cite{bao:2003,arlett:2005}. For example, the force
measured by Radmacher et al. as the protein lysozyme metabolizes substrate is
approximately $50$pN~\cite{radmacher:1994}.

\begin{figure}[htb]
  \begin{center}
    \includegraphics[height=1.5in]{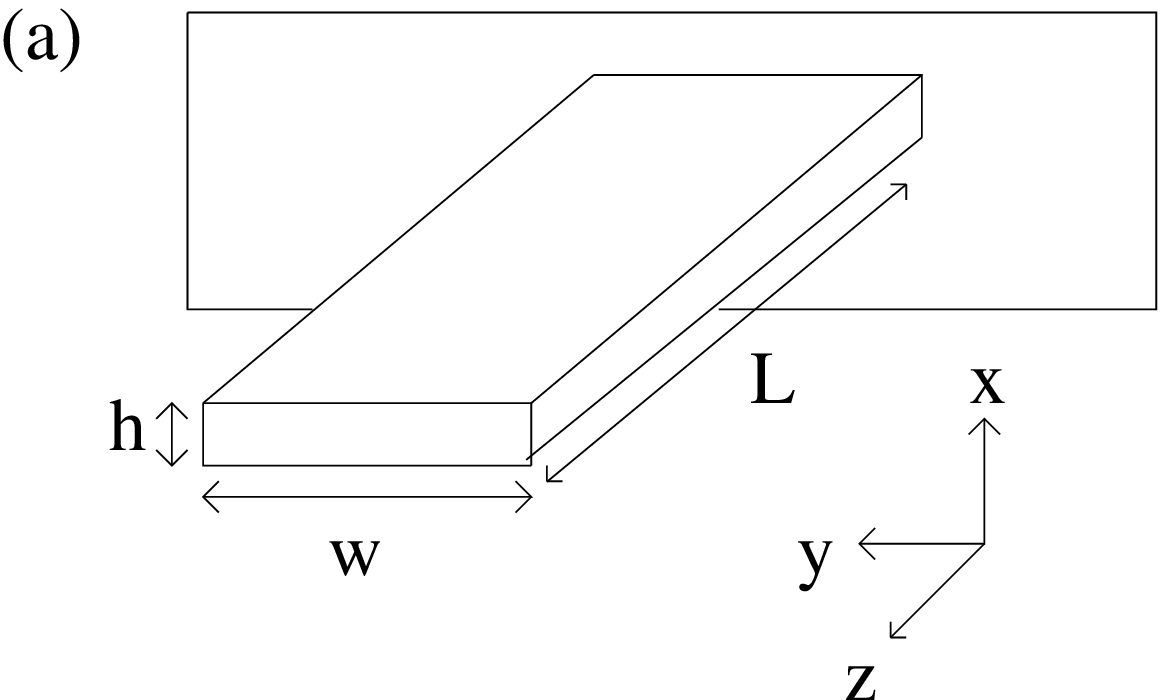}
    \includegraphics[width=2.0in]{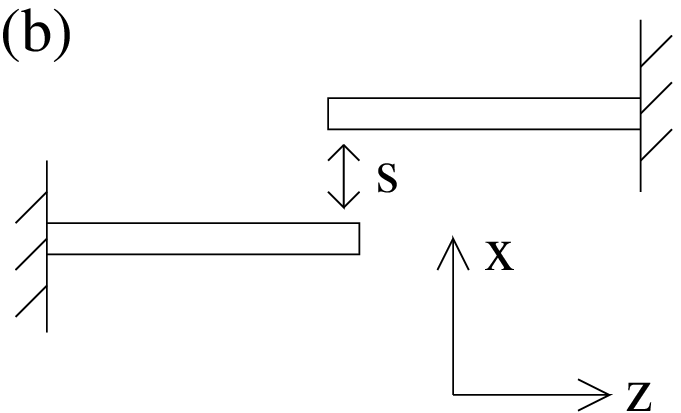}
  \end{center}
  \caption{(a)~The convention used in defining the cantilever geometry, not drawn
  to scale.  Actual aspect ratios are $L/h = 98.5$, $w/h = 14.5$, and
  $L/w = 6.8$. Specific properties of the
  cantilevers explored here are given in Table~\ref{table:cantilever}.
  (b)~The configuration of two adjacent AFM cantilevers (not drawn to scale). In the
  absence of a tethered biomolecule between the cantilevers their motion is correlated
  due to the induced fluid motion.}
  \label{fig:setup}
\end{figure}

However, such a measurement technique can be significantly
improved through the detection of \textit{correlations} in the
displacements of the two adjacent cantilevers, for example see
Fig.~\ref{fig:setup}(b). The motion of one cantilever will cause
fluid motion that will move the adjacent cantilever and vice
versa. The Brownian force on each cantilever is uncorrelated and
does not contribute to a cross-correlation measurement.  The
only contribution comes from the correlations due to the fluid
which will be significantly reduced in magnitude when compared to
the Brownian force felt by a single cantilever. For example,
femtonewton forces have been measured from the correlated motion
of micron scale spheres held in closely spaced optical
traps~\cite{meiners:1999,meiners:2000}. Now consider tethering a
biomolecule between the two cantilevers in
Fig.~\ref{fig:setup}(b). In this case, the motion of the two
cantilevers will be correlated by the fluid as well as by the
dynamics of the linking target biomolecule. Measurements of the
correlations due to the biomolecule could be used to detect the
presence of the biomolecule and to probe its biomolecular dynamics
in real time.

Prior to the development of such experimental methods the fluidic
coupling between adjacent AFM cantilevers must be understood in
order to build, design and interpret experiments where the noise
from fluid induced correlations is small compared to the forces
caused by the dynamics of the tethered biomolecule. The magnitude
of the noise induced by fluid correlations in an array of
cantilevers is complex and depends upon many factors including
cantilever geometry and array configuration.  In this paper we
explore the fluid induced correlations between two readily
available atomic force microscopes in an experimentally motivated
opposing configuration. This will provide a baseline understanding
of their stochastic dynamics in the absence of a target
biomolecule which is essential to the success of future
experiments.

Although the cantilever motion is driven by molecular collisions
the equations necessary to describe these dynamics are those of
classical continuum mechanics. The Knudsen number, Kn, represents
the ratio of the mean free path between collisions $\lambda$ of a
fluid molecule to a characteristic length $L$ in the system. In
the limit of Kn$\rightarrow 0$ the continuum hypothesis is valid,
however it has been found experimentally to be a good
approximation for Kn $\lesssim 1 \times
10^{-3}$~\cite{schaaf:1961,karn:2001}. For larger values of Kn the
continuum hypothesis fails with such consequences as the violation
of the no-slip boundary condition at fluid solid interfaces. The relevant 
length scale $L$ for very long cantilevers is 
the cantilever half-width $w/2$ from Table~\ref{table:cantilever}
and $\lambda$ is approximately the diameter of a water molecule, yielding Kn
$= \lambda/L \approx 10^{-5}$. When an array configuration is considered
a characteristic length may also be chosen as the spacing between the cantilevers.
For the smallest separation investigated here, $s = 200$nm, Kn $\approx 10^{-4}$.
Since Kn is small in both situations we will assume
that both the continuum and no-slip hypotheses are valid.

To describe the cantilever dynamics the equations of
elasticity~\cite{landau:1959} must be coupled with the equations
of fluid dynamics~\cite{panton:2005}. The fluid coupling of an
array of cantilevers is quite complex, and as a result it is
useful to discuss the fluid equations in more detail. The
Navier-Stokes equations for incompressible fluids are,
\begin{eqnarray}
\beta~\frac{\partial \vec{u}}{\partial t} + R~\vec{u} \dotprod
\vec{\nabla} \vec{u} &=& -\vec{\nabla} p + \nabla^2 \vec{u}, \label{eq:nd1} \\
\vec{\nabla} \dotprod \vec{u} &=& 0, \label{eq:nd2}
\end{eqnarray}
where $\vec{u}=(u,v,w)$ is the fluid velocity and $p$ is the
pressure. The equations are nondimensionalized using the
cantilever half-width $w/2$, the maximum cantilever oscillation
velocity $U$, and the inverse cantilever oscillation frequency
$\omega^{-1}$ as the characteristic length, velocity, and time
scales, respectively. On all material boundaries we impose the
no-slip fluid boundary condition. The frequency parameter $\beta =
w^2\omega/4 \nu$ is a frequency based Reynolds number and
represents the ratio of local inertial forces to viscous forces
where $\nu$ is the kinematic viscosity of the liquid. The Reynolds
number $R= U w/2 \nu$ represents the ratio of convective inertial
forces to viscous forces. For the high frequency and low amplitude
oscillations of interest here $R \ll 1$ and $\beta \sim 1$
resulting in the unsteady Stokes equations.

An often used approximation for the flow field generated by an
oscillating atomic force microscope is that of an infinite
cylinder performing transverse oscillations~\cite{sader:1998}. The
fluid dynamics describing the flow field generated can be
described in terms of viscous and potential
contributions~\cite{rosenhead:1963}. The potential component
reacts instantaneously and depends only upon the position of the
cylinder and is out of phase with the cylinder motion. The viscous
component is characterized by the diffusion of momentum from the
cylinder surface with diffusion constant $\nu$. The phase of the
viscous component depends in a complicated manner upon the
position of the cylinder. The interplay between the potential and
viscous components of the flow field can lead to complex dynamics.

An estimate of the length scale over which viscous effects
diffuse during a single cantilever oscillation is given by the
Stokes length,
\begin{equation}
\delta_s \approx \left( \frac{\nu}{\omega_f} \right)^{1/2} = \frac{w}{2} \beta^{-1/2},
\end{equation}
where $\omega_f$ is the resonant frequency for the beam in fluid.
For microscale systems this length scale can become quite large,
for example a cantilever oscillating at 50 kHz in water yields a
Stoke's length on the order of 2 $\mu$m. For very small cantilever
separations, as of interest here, the two cantilevers will be
immersed in each others Stoke's layers and their dynamics will be
determined by the interactions of the viscous and potential
responses. It is important to note that although the fluid
dynamics of simple oscillating objects such as cylinders, spheres,
and ellipsoids are well known the coupled fluid dynamics of an
array of such objects is poorly understood.

The dynamics of a single micron scale cantilever in vacuum can be
described using the equipartition theorem for systems in thermal
equilibrium~\cite{butt:1995}. This approach has been extended to
include the damping effects of a viscous fluid~\cite{sader:1998}.
In this case, the cantilever was assumed to be long and thin and,
as a result, the stochastic dynamics were described by coupling
the classical elasticity equations to the flow field caused by an
oscillating infinite cylinder. A further simplification
in~\cite{sader:1998} is the assumption that the noise force is
frequency independent which is not strictly
justified~\cite{paul:2004,paul:2006}.

A recently proposed thermodynamic approach uses the
fluctuation-dissipation theorem to predict the stochastic
cantilever dynamics through \textit{deterministic} calculations of
the coupled fluid-solid equations~\cite{paul:2004,paul:2006}. Only
a brief overview of the method is given here, see
Refs.~\cite{paul:2004,paul:2006} for a detailed discussion.  A
convenient way to use this approach is to calculate the
deterministic dynamics of the cantilevers for the case where one
cantilever is exposed to a small step force $f(t)$ given by,
\begin{equation}
f(t)=\left\{
\begin{array}
[c]{cc}%
F_1 & \text{for }t<0\\
0 & \text{for }t\ge0.
\end{array}
\right. \label{eq:step_force}
\end{equation}
In our simulations the step force is applied to the lower
cantilever in Fig.~\ref{fig:setup}(b). Upon removal of the force
the lower cantilever returns to equilibrium through underdamped
oscillations given by the position of the cantilever tip $X_1(t)$.
The fluid motion generated as the lower cantilever returns to
equilibrium causes deflections in the tip of the upper cantilever
given by $X_2(t)$. The auto and cross-correlations of the
equilibrium fluctuations in cantilever displacement are then given
by,
\begin{equation}
\left\langle x_1(0) x_j(t)\right\rangle=\frac{k_{B}T}{F_1}
X_j(t), \label{eq:autocorr}
\end{equation}
where $X_j(t)$ is the deflection of the $j^{th}$ cantilever, $k_B$
is Boltzmann's constant, $T$ is the absolute temperature, and
$\left< \right>$ denotes an equilibrium average in the absence
of the step force $f$. For clarity, a capital $X(t)$ indicates the
deterministic cantilever deflection and a small $x(t)$ indicates
the stochastic deflection. When $j = 1$ Eq.~(\ref{eq:autocorr})
yields the autocorrelation, and for $j = 2$ it yields the cross
correlation. The spectral properties of the correlations are given
by the Fourier transform of the auto and cross-correlations to
yield the noise spectra, $G_{11}(\omega)$ and $G_{12}(\omega)$.

\section{The stochastic dynamics of an array of atomic force microscope cantilevers}
To explore the stochastic dynamics of two adjacent cantilevers we
have performed time-dependent, three dimensional finite element
simulations of the governing deterministic fluid-solid equations
(algorithm described elsewhere~\cite{cfdrc,yang:1994}). The 
AFM cantilevers are micron scale with the simple beam
geometry shown in Fig.~\ref{fig:setup}(a). The physical properties
of a cantilever are summarized in Table~\ref{table:cantilever}.
The stochastic dynamics of a single cantilever with this geometry
in fluid has been explored both experimentally~\cite{chon:1999}
and theoretically~\cite{paul:2004,paul:2006}. The characteristic
quantities to describe its motion are given here in
Table~\ref{table:fluid}~\cite{paul:2006}. Most atomic force
microscope cantilevers have a probe on the distal end that
interacts with the sample of interest~\cite{garcia:2002}. However,
in this work we are interested in characterizing the fluid
coupling between two stochastic cantilevers and, as a result, we
explore the simpler case of two rectangular beams without attached
probes. Experimental features such as a probe could be included if
desired.
\begin{figure}[htb]
  \begin{center}
    \includegraphics[width=3.25in]{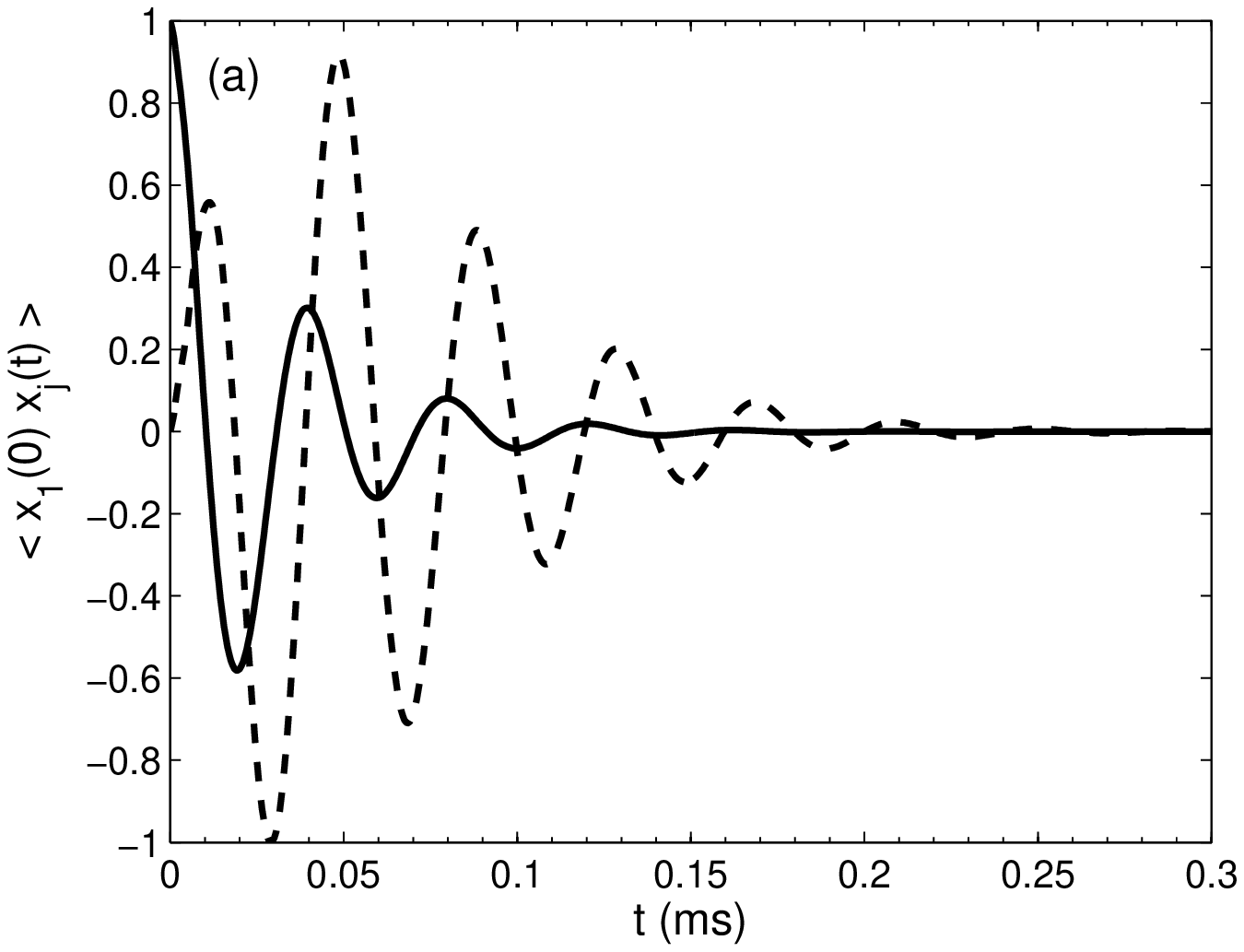}
    \includegraphics[width=3.25in]{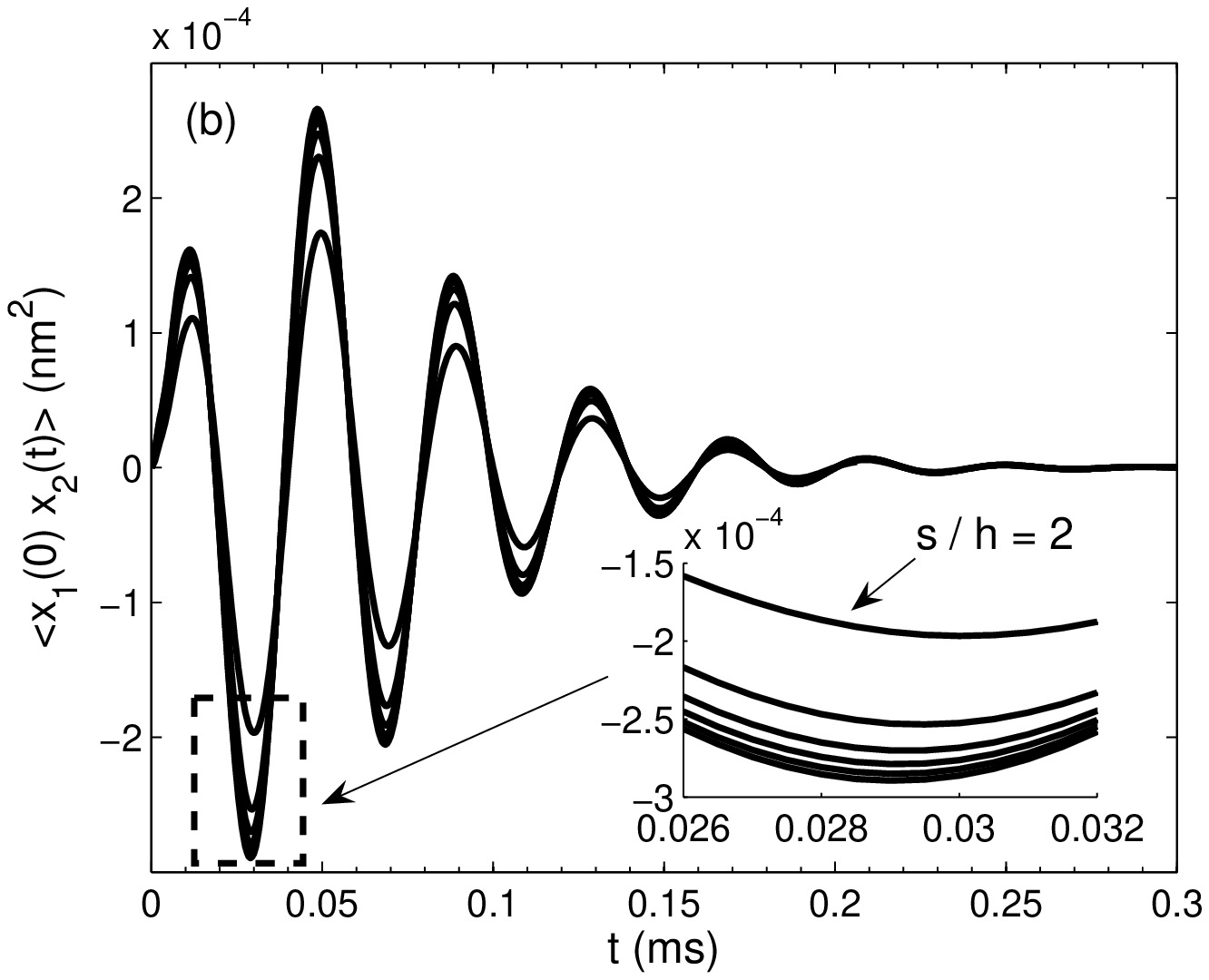}
  \end{center}
  \caption{(a) The autocorrelation (solid line) and cross-correlation (dashed line) of
  the equilibrium fluctuations in cantilever displacement for $s/h = 0.1$.
  The two functions have been normalized to highlight the phase difference in the
  dynamics. The normalization used is $3 \times 10^{-3}$nm$^2$ for the
  autocorrelation and $2.9 \times 10^{-4}$nm$^2$ for the cross-correlation. (b) Cross-correlation of
  the equilibrium fluctuations in cantilever displacement over the range of cantilever
  separations $s/h = 0.1, 0.3, 0.5, 0.7, 1, 2$. (inset) Detailed view of the
  largest magnitude cross-correlations illustrating a decreasing magnitude with
  cantilever separation.}
  \label{fig:pascorr}
\end{figure}
\begin{table}[tbp]
\begin{center}
\begin{tabular}
[c]{l@{\hspace{0.5cm}}l@{\hspace{0.5cm}}l@{\hspace{0.5cm}}l@{\hspace{0.5cm}}
l@{\hspace{0.5cm}}l@{\hspace{0.5cm}}l@{\hspace{0.5cm}}l}
$L$ & $w$ & $h$ & $k$ & $\omega_0$
\\ \hline \hline \\
197$\mu$m & 29$\mu$m & 2.0$\mu$m & 1.32 N/m & $452 \times 10^3$ rad/s
\end{tabular}
\end{center}
\caption{Summary of the cantilever geometry and properties. The
geometry is given by the cantilever length $L$, width $w$, and
height $h$. The cantilever spring constant is $k$ and the resonant
frequency in vacuum is $\omega_0$. The cantilevers are composed of silicon with
density $\rho_c$ = 2320 kg/$\text{m}^3$, and Youngs modulus E =
1.74 $\times$ $10^{11}$ N/$\text{m}^2$. The cantilevers are
immersed in water with density $\rho_l$ = 997 kg/$\text{m}^3$ and
dynamic viscosity $\eta$ = 8.59 $\times$ $10^{-4}$ kg/m-s.}
\label{table:cantilever}
\end{table}
\begin{table}[tbp]
\begin{center}
\begin{tabular}
[c]{l@{\hspace{0.5cm}}l@{\hspace{0.5cm}}l@{\hspace{0.5cm}}l@{\hspace{0.5cm}}l@{\hspace{0.5cm}}l}
 $m_f/m_e$ & $\omega_f/\omega_0$ & $Q$ & $\beta$ & $R$ \\ \hline \hline \\
 8.2       & 0.35                & 3.0 & 39.0    & $3.0 \times 10^{-4}$
\end{tabular}
\end{center}
\caption{Summary of the cantilever dynamics in fluid as determined
from numerical simulations in Ref~\cite{paul:2006}: the fluid
loaded mass of the cantilever $m_f$, the equivalent cantilever
mass $m_e$, resonant frequency in vacuum $\omega_0$, resonant
frequency in fluid $\omega_f$, the quality $Q$, the frequency
parameter $\beta$, and the Reynolds number $R$. The values of
$m_f$, $\omega_f$, and $Q$ have been determined by fitting the
cantilever response to a simple harmonic oscillator.}
\label{table:fluid}
\end{table}

A series of numerical experiments are performed for a range of
cantilever separations $s$, given by $0.1 \le s/h \le 2$ where the
separation is measured as the distance between the cantilever tips
at zero deflection. The minimum cantilever separation investigated
here, $s = 200$nm, is a result of computational constraints given
by the numerical method used. The actual cantilever separation in
a two-cantilever experiment will depend upon the size and geometry
of the cantilever probes~\cite{garcia:2002} and the particular
biofunctionalization approach used to tether the
biomolecule~\cite{arlett:2005,solomon:2006}. Considering readily
available cantilever tips and biofunctionalization protocols the
range of cantilever separations explored spans what would be
expected in experiment~\cite{nanoscience,arlett:2005}.

The numerical simulations yield $X_1(t)$ and $X_2(t)$ and the auto
and cross-correlations of the equilibrium fluctuations in
cantilever displacement are found using Eq.~(\ref{eq:autocorr}).
The auto and cross-correlations are shown in
Fig.~\ref{fig:pascorr} as a function of cantilever separation. The
autocorrelation $\left< x_1(0) x_1(t) \right>$, normalized by its
maximum value, is given by the solid line in
Fig.~\ref{fig:pascorr}(a). The autocorrelation did not depend upon
the presence of the adjacent cantilever for all cantilever
separations tested and the curve shown is representative for all
simulations. The cross-correlation for $s/h = 0.1$, normalized by
its maximum value, is also plotted in Fig.~\ref{fig:pascorr}(a) to
clearly illustrate the phase relationship between the auto and
cross-correlations. The cross-correlations in the equilibrium
fluctuations $\left< x_1(0) x_2(t) \right>$ are shown in
Fig.~\ref{fig:pascorr}(b). As expected, the magnitude of the
cross-correlation decreases as the separation between the two
cantilevers is increased (see inset of Fig.~\ref{fig:pascorr}(b)).

\begin{figure}
\begin{center}
\includegraphics[width=2.5in]{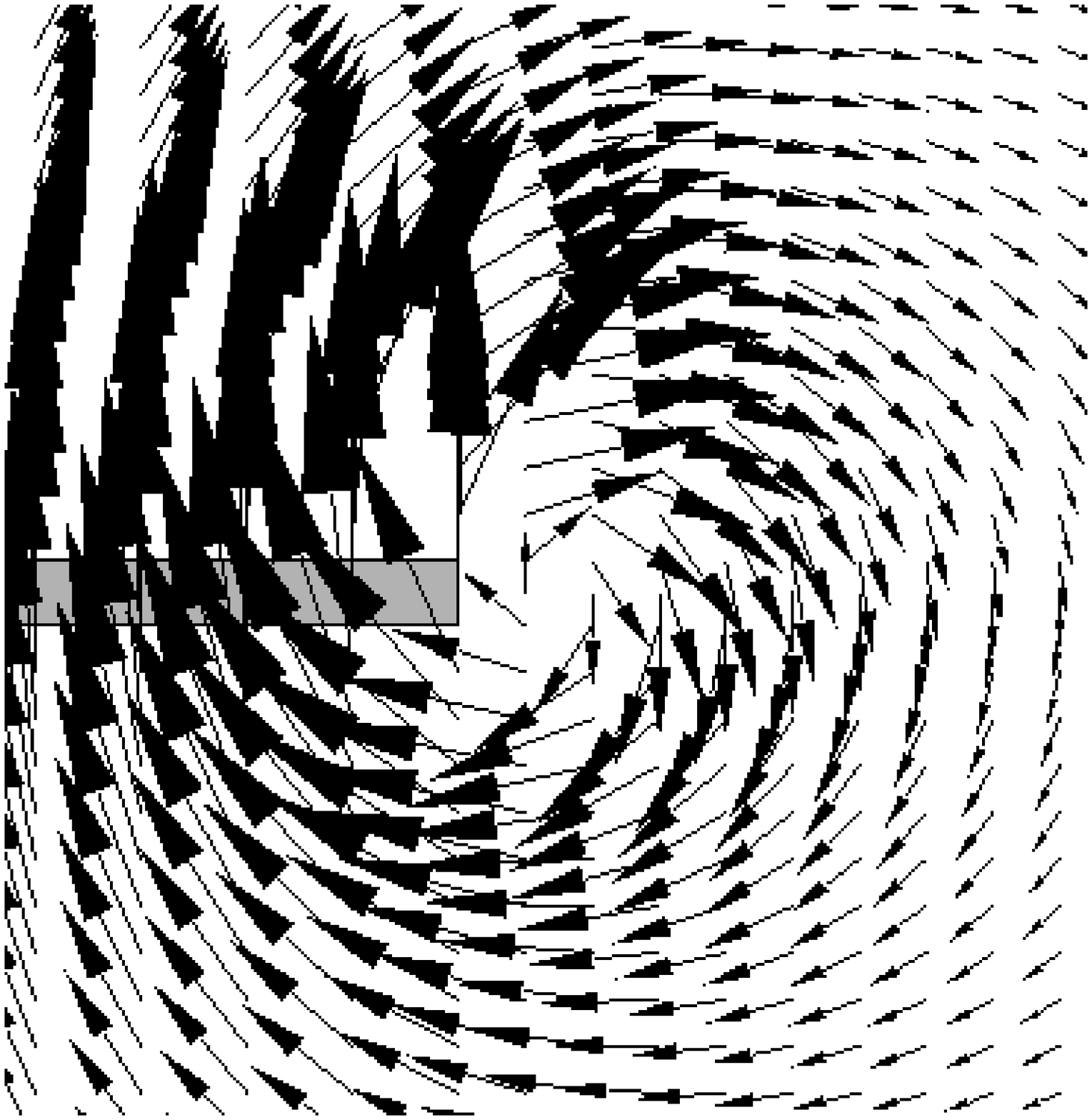} \\
 \vspace{0.1in}
\includegraphics[width=3in]{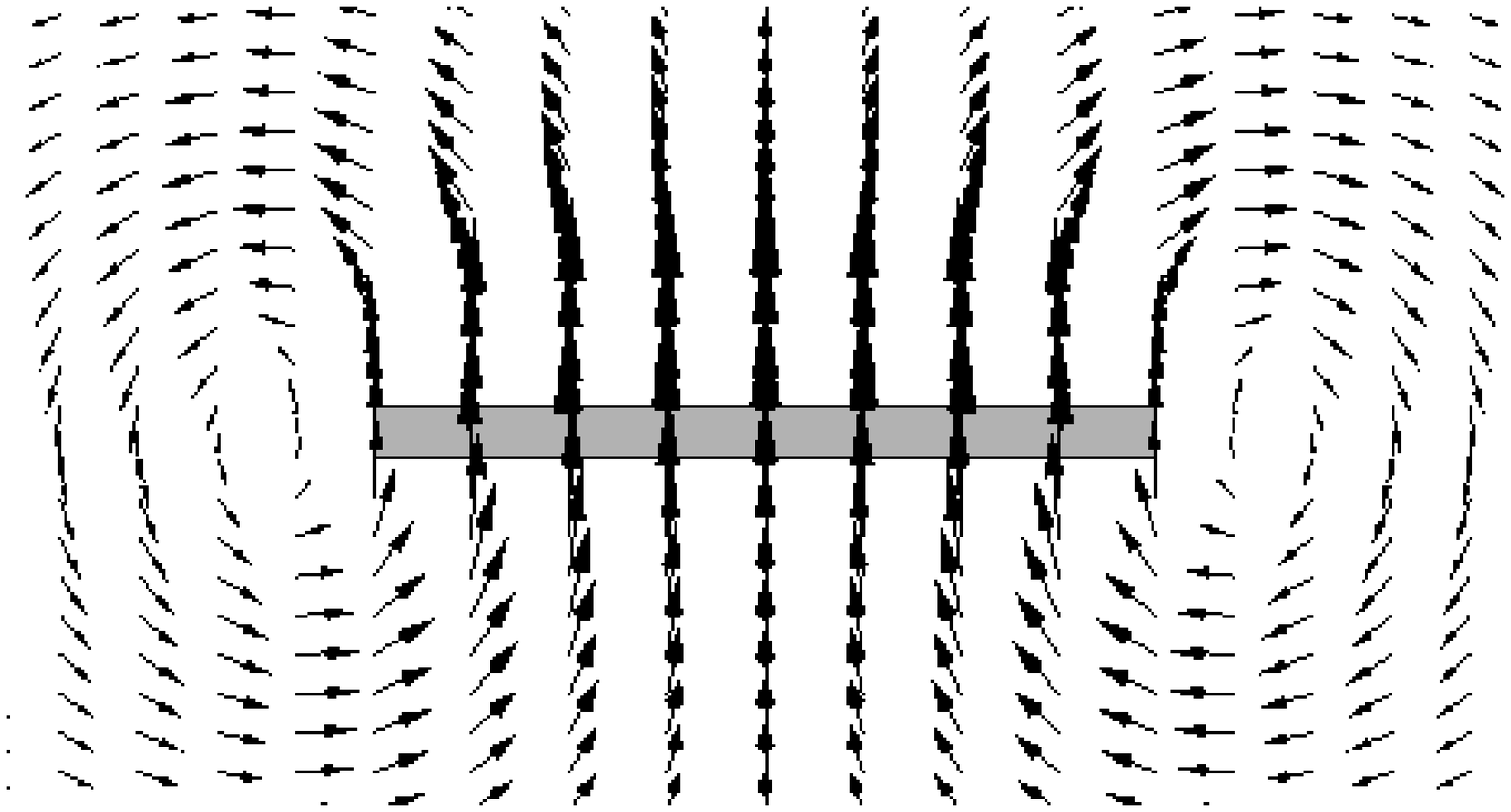}
\caption{The three-dimensional flow field near the tip (distal
end) of an oscillating atomic force microscope. The flow field is
a snap shot in time at the instant where the cantilever is at its
maximum velocity, $t^*=9.0\mu$s. (top) Flow field over the
cantilever tip in the the $x-z$ plane. The base of the cantilever
is to the far left of the figure and is not shown. (bottom) Flow
field around the sides of the cantilever in the $x-y$ plane. See
Fig.~\ref{fig:setup} for coordinate axis definitions. In both
figures the largest arrow indicates a fluid velocity of $U =
2.5$ mm/s.} \label{fig:flowfield}
\end{center}
\end{figure}

\begin{figure}
\begin{center}
\includegraphics[width=2.5in]{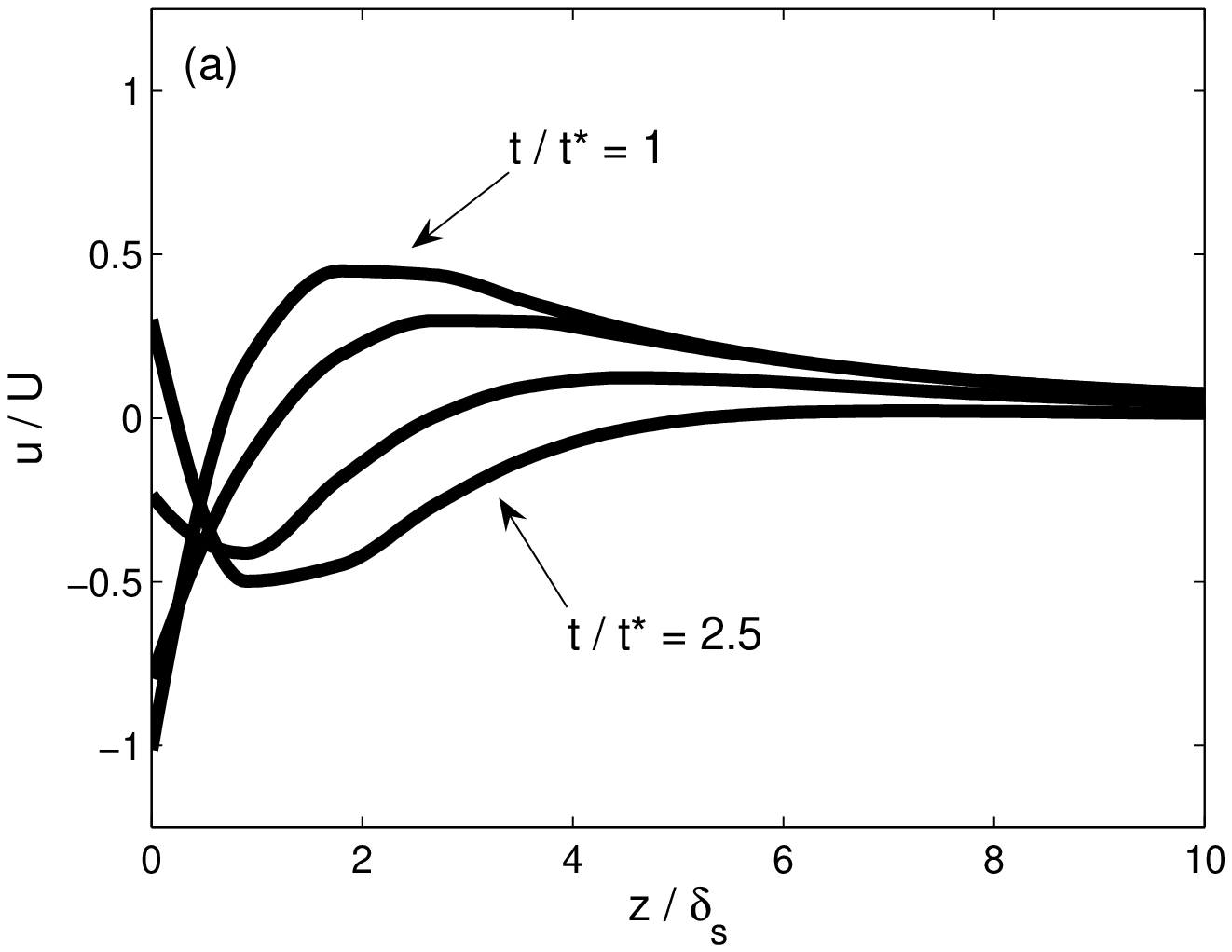}
\includegraphics[width=2.5in]{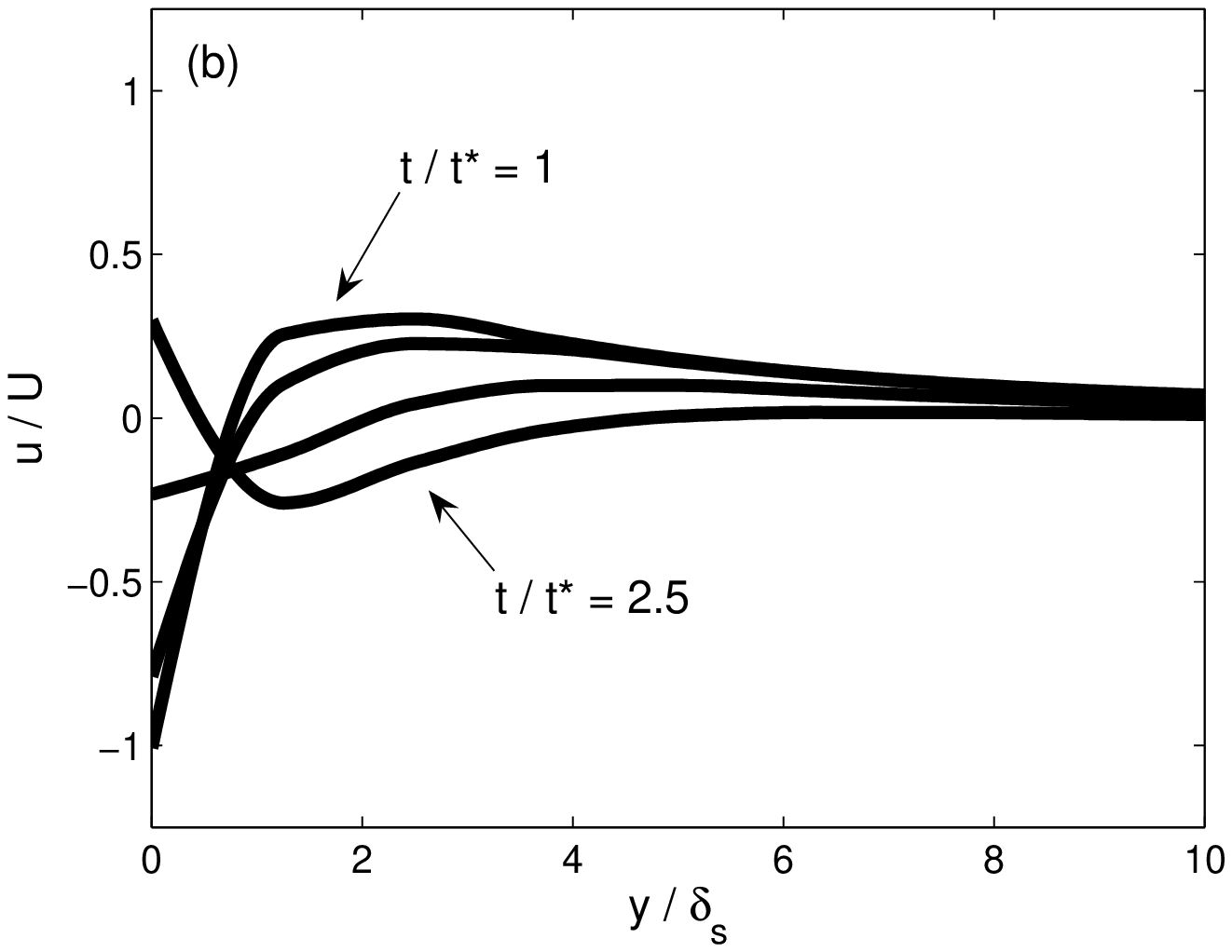}
\caption{(a) The variation in the x-component of the fluid
velocity $u$ as a function of distance in the $z$-direction
measured from the tip of the cantilever at various times during
the cantilevers oscillations. (b) The variation in the x-component
of the fluid velocity $u$ as a function of distance in the
$y$-direction measured from the tip of the cantilever at various
times during the cantilevers oscillations. In each plot curves are
shown for $t/t^*=1,1.5,2,2.5$, where $t/t^*=1$ and 2.5 are labeled
and the other two are in sequence.} \label{fig:fluidvel}
\end{center}
\end{figure}

It is useful to look closer at the deterministic flow field caused
by an oscillating cantilever that would yield the 
autocorrelations through Eq.~\ref{eq:autocorr}. Figure~\ref{fig:flowfield} illustrates two different
cross-sections of the flow field at $1.5 t^*$ where $t^*$ is the
time at which the cantilever reaches its maximum velocity. This
corresponds to a cantilever tip velocity of $U = 2.5$ mm/s
which occurs at a time $t^* = 9.0 \mu$s, where $t^*/t_p \approx
0.22$ and $t_p$ is the time required for the cantilever to
complete its first oscillation. The magnitude of the fluid
velocity is less than $1\%$ of the maximum value after a time of
$t \approx 20 t^*$. The numerical simulation used to generate the
flow field of Fig.~\ref{fig:flowfield} was for only a single
cantilever in fluid to clearly illustrate the three-dimensional
nature of the flow field around the tip of an oscillating
cantilever. A closer inspection of the flow field reveals that
finite fluid velocities extend further in the $z$-direction than
in the $y$-direction as shown in panels~(a) and~(b) of
Fig~\ref{fig:fluidvel}, respectively. In other words, the fluid
flow over the cantilever tip is greater than the fluid flow around
the sides of the cantilever. This suggests that an elastic object
located to the side of the cantilever (in the $y$-direction) will
exhibit less fluid coupling than an object placed at the same
distance off of the tip of the cantilever (in the $z$-direction).

\begin{figure}[htb]
  \begin{center}
    \includegraphics[width=3.25in]{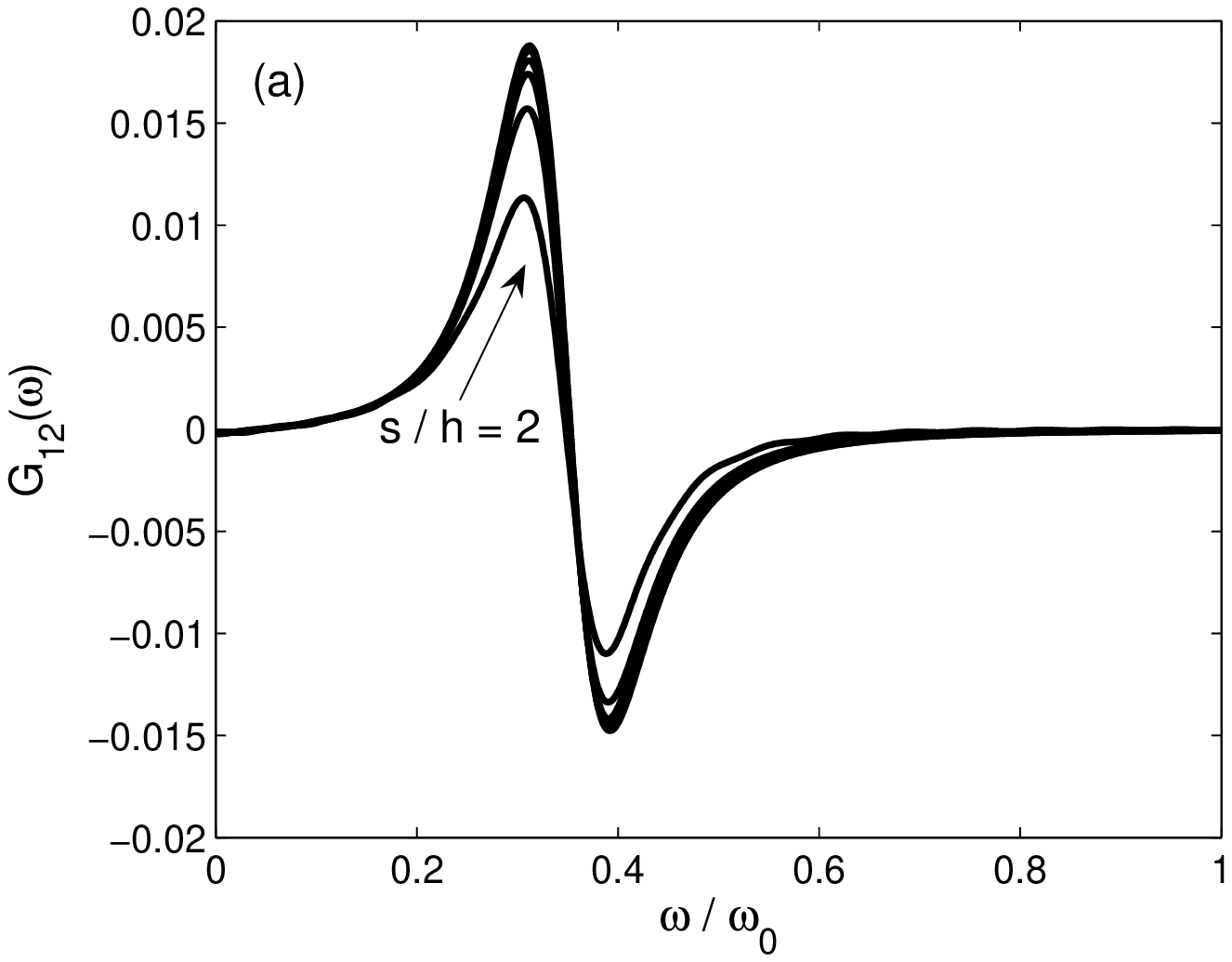}
    \includegraphics[width=3.25in]{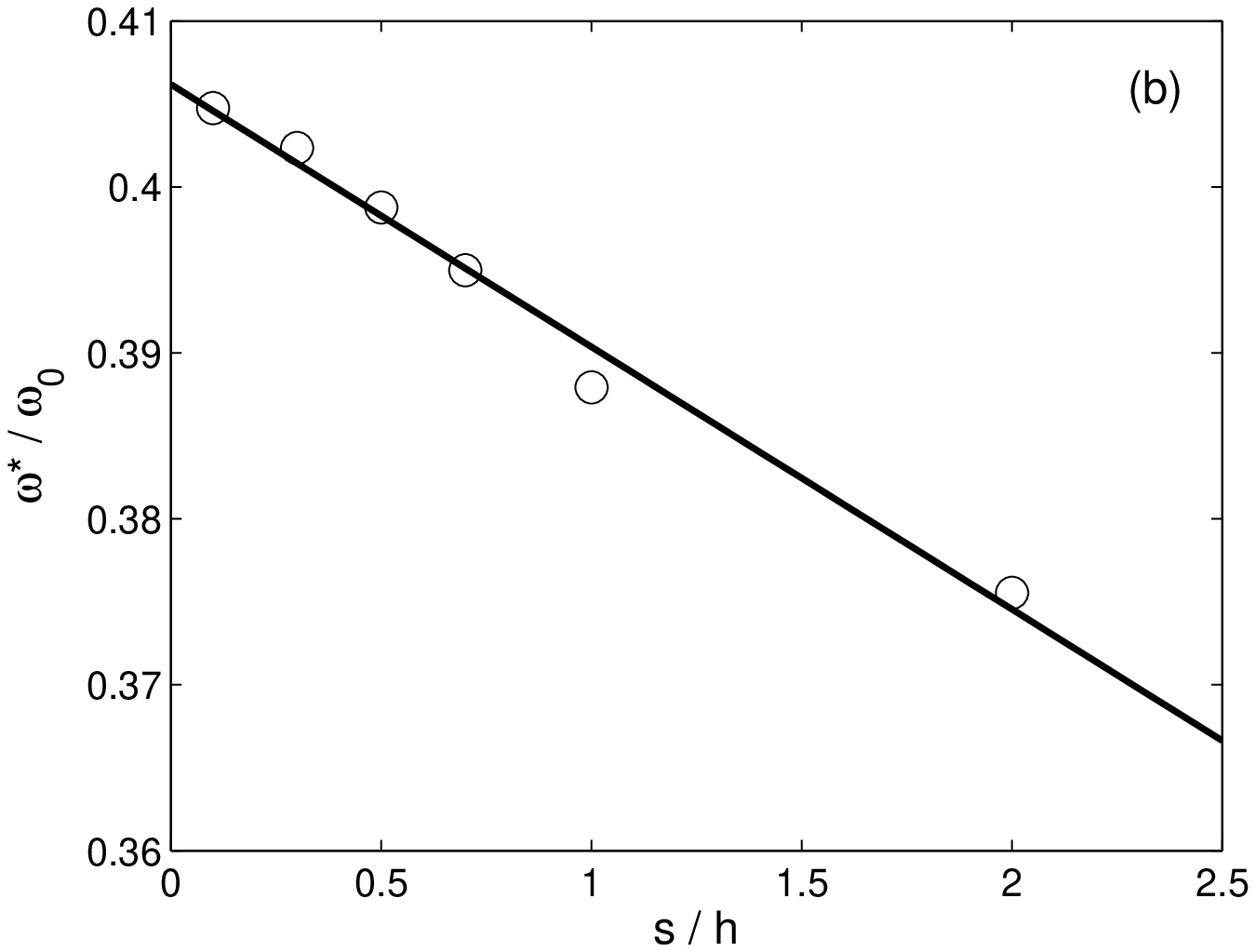}
  \end{center}
  \caption{(a) The noise spectrum $G_{12}(\omega)$ for the cantilever array,
  normalized by the same factor as $G_{11}(\omega)$ in~\cite{paul:2006}. Results
  are shown for cantilever separations of $s/h = 0.1, 0.3, 0.5, 0.7, 1, 2$.
  The curve for $s/h=2$ is labelled and the others are in sequential order.
  (b) The frequency at which the spectral response crosses zero $\omega^*$ as a
  function of separation. The solid line is a linear curve fit given by
  $\omega^* / \omega_0 = -1.58 \times 10^{-2}(s/h) + 0.406$.}
  \label{fig:pasnoise}
\end{figure}

Fourier transforms of the cross-correlations yield the noise
spectra which are shown in Fig.~\ref{fig:pasnoise}(a) as a
function of cantilever separation. It is interesting to note that
there is a particular frequency $\omega^*$ where the noise spectra
vanishes, $G_{12}(\omega^*)=0$. This knowledge could lead to
experimental protocols to minimize the fluid correlated noise. To
illustrate this further we plot the behavior of $\omega^*$ in
Fig.~\ref{fig:pasnoise}(b). The solid line is a linear fit to the
data illustrating a steady decrease in $\omega^*$ with increasing
separation. The reduced frequencies for larger separations are a
result of the finite time at which the viscous effects propagate.
\begin{figure}[htb]
  \begin{center}
    \includegraphics[height=2.4in]{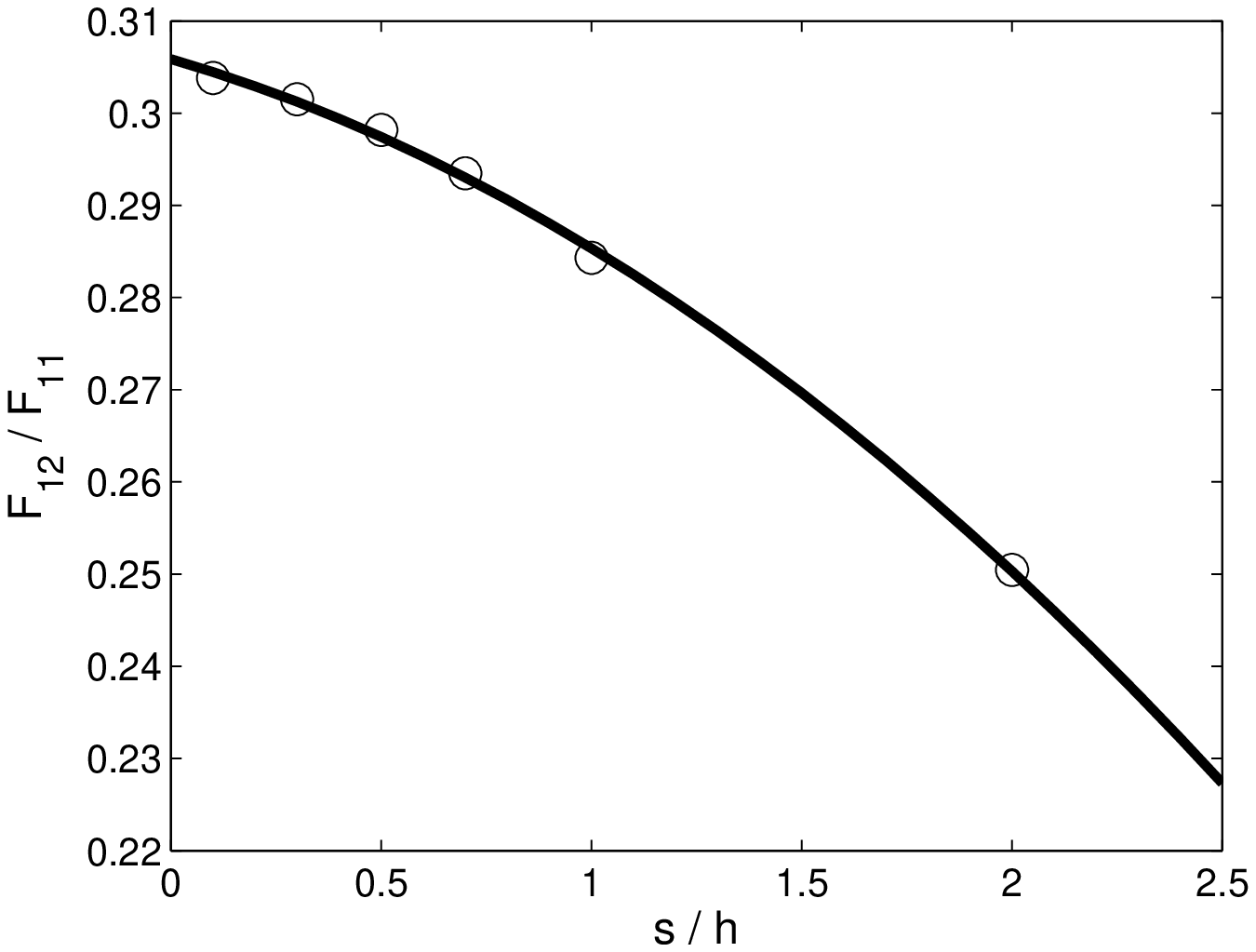}
  \end{center}
  \caption{Force sensitivity $F_{12}$ as a function of AFM cantilever separation.
  Forces have been normalized by the Brownian force on a single cantilever
  ($F_{11}$ = 74 pN). The solid line is a quadratic curve fit to
  the data and is given by
  $F_{12} / F_{11} = -7.3 \times 10^{-3} (s/h)^2 - 1.33 \times 10^{-2} (s/h) + 0.306$.}
  \label{fig:fsens}
\end{figure}

Using these results we can characterize the force sensitivity and
time resolution of a correlated measurement technique using the
cantilever array. An estimate of the force sensitivity can be
found using the auto and cross-correlation
functions~\cite{paul:2006},
\begin{eqnarray}
  \label{fsens}
  F_{11} &=& k |\langle x_1(0) x_1(t) \rangle|_{max}^{1/2}, \\
  \label{fsens2}
  F_{12} &=& k |\langle x_1(0) x_2(t) \rangle|_{max}^{1/2}.
\end{eqnarray}
In our notation, $F_{11}$ represents the magnitude of the
stochastic Brownian force acting on a single cantilever. As shown
by the solid line in Fig.~\ref{fig:pascorr}(a) the maximum value
of the autocorrelation occurs at time $t=0$ which is the
root-mean-squared deflection of the stochastic motion. Using
the equipartition theorem, this is given by the simple
expression~\cite{albrecht:1987},
\begin{equation}
 \left< x^2 \right>^{1/2} = \sqrt {\frac{k_B T}{k}},
\label{eq:equipartion}
\end{equation}
which yields $0.56 \AA$. The magnitude of the thermally driven
oscillations are much smaller than the thickness of the
cantilever, $\left< x^2 \right>^{1/2}/h = 2.8 \times 10^{-5}$.  In
this case, the Brownian force on an individual cantilever is
$F_{11} = \left( k_B T k \right)^{1/2} = 74$pN.

In a cross-correlation measurement between two cantilevers the
Brownian noise felt by the two individual cantilevers is
uncorrelated and does not contribute. This leaves only the
correlations due to the fluidic induced correlations. $F_{12}$
represents the approximate magnitude of the force due to these
hydrodynamic correlations. From Fig.~\ref{fig:pascorr}(b) for the
case with the closest separation, $s/h=0.1$, the maximum magnitude
of the cross-correlation $\left| \langle x_1(0) x_2(t) \rangle
\right|_{max} \approx 2.9 \times 10^{-4} $nm$^2$ which yields a
root-mean-squared displacement of $0.17 \AA$. Using
Eq.~(\ref{fsens2}) this yields a force sensitivity of approximately
22 pN, an improvement of better than three fold over the force
noise acting on a single cantilever. The variation in $F_{12}$ with
cantilever separation is shown in Fig.~\ref{fig:fsens}. The force
due to fluid correlations decreases with separation and the solid
line represents a quadratic polynomial curve fit to the data. The
reduction in the noise is quite gradual over the separations of
interest. The noise reduction at the largest separation $s=2\mu$m
is a 4-fold improvement over a single cantilever.

An estimate of the time resolution possible is given by the time
it takes the cantilever to complete an oscillation at its resonant
frequency. Using the parameters obtained from the simple harmonic
oscillator curve fit for a single cantilever in fluid, this time
scale is estimated to be $\tau \approx 2 \pi / \omega_f = 39 \mu$s
which yields a measurement frequency of 25 kHz. Therefore the
cantilever array investigated here can measure forces on the order
of 10's of pN with kHz frequency resolution which would make
possible the real time measurements of many interesting molecular
interactions~\cite{radmacher:1994,bao:2003}.

As mentioned, the length over which viscous effects act can be
quite large for microscale systems. To ensure that the simulation
boundaries do not significantly affect the results a series of
numerical tests were performed using varying numerical domain
sizes. The size of the numerical domain to be used for analysis
was chosen such that the magnitude of the fluid velocity went
gradually to zero at the walls. Specifically, the distance between
the cantilever tip and any numerical bounding wall is chosen such
that the velocity field at $t/t^*=1$ gradually decreases to a
vlaue less than $1\%$ of the maximum velocity before reaching a
solid wall. Numerical results indicate that the distance from the
cantilever tip to the walls should be at least $24 \delta_s$ in
both $y$ and $z$, and $12 \delta_s$ in $x$. The numerical domain
was chosen to be larger than these constraints in each direction
for all numerical simulations presented here. Results from
simulations where the numerical domain was smaller than these
values yielded cantilever dynamics with lower values of $Q$ and
$\omega_f$ in agreement with the predicted increase in dissipation
for a cantilever oscillating near a
wall~\cite{green:2005,clarke:2005}.

\section{Conclusions}
The correlated stochastic dynamics of an array of two AFM
cantilevers have been obtained using a thermodynamic approach with
deterministic numerical calculations. Measurements of the
correlations in cantilever displacement yield force sensitivities
that are improved by a factor of 3 to 4 over the range of interest
for cantilever separations that may be useful for single molecule
experiments. The complex flow field around the tip of an
oscillating cantilever suggests that the configuration of the
cantilever array is an important determining factor in the
resulting cantilever correlations. The approach used here is quite
general and can be extended to include more complex geometries and
array configurations that are motivated by experiment. Theoretical
calculations, such as these, will provide important insight
necessary for the interpretation and design of future micro and
nanoscale technologies that exploit inherent thermal fluctuations.

\section{Acknowledgments}
This research has been partially supported by DARPA/MTO Simbiosys
under grant F49620-02-1-0085 and an ASPIRES grant from Virginia
Tech. We gratefully acknowledge extensive interactions with S.
Quake, M.C. Cross, and the Caltech BioNEMS effort (M.L. Roukes,
PI).
\bibstyle{prsty}
\bibliography{mtc2005}
\end{document}